\documentclass[12pt]{extarticle}
\usepackage{amsmath, amsthm, amssymb, hyperref, color}
\usepackage{graphicx}
\usepackage[all]{xypic}
\usepackage{verbatim}
\usepackage{tikz}

\oddsidemargin \evensidemargin
\usepackage{color}
\newif\ifincludeprevious

\newtheorem{theorem}{Theorem}
\numberwithin{theorem}{section}

\newtheorem{remark}[theorem]{Remark}

\newenvironment{proof_of}[1]{\noindent {\bf Proof of #1:}
	\hspace*{1mm}}{\hspace*{\fill} $\Box$ }

\newcommand{\tnull}{\text{null}}

\usepackage{listings}
\date{}
 
\title{\textbf{Computing Theta Functions with Julia}}

\author{Daniele Agostini and Lynn Chua}

\begin{document}

\maketitle

\begin{abstract}
We present a new package \texttt{Theta.jl} for computing the Riemann theta function. It is implemented in Julia and offers accurate numerical evaluation of theta functions with characteristics and their derivatives of arbitrary order. Our package is optimized for multiple evaluations of theta functions for the same Riemann matrix, in small dimensions. As an application, we report on experimental approaches to the Schottky problem in genus five.
\end{abstract}

\section{Introduction} \label{section:intro}
The \emph{Riemann theta function} is the holomorphic function
\begin{equation}\label{eqn:theta}
\theta \colon \mathbb{C}^g \times \mathbb{H}_g \to \mathbb{C}\,, \qquad \theta(z,\tau) = \sum_{n\in \mathbb{Z}^g} \mathbf{e}\left( \frac{1}{2}n^t \tau n + n^t z \right)
\end{equation}
where $\mathbf{e}(x)= e^{2\pi i x}$ and $\mathbb{H}_g$ is the \emph{Siegel
upper-half space}, which consists of all complex symmetric $g\times g$
matrices with positive definite imaginary part. Theta functions occupy a central role throughout mathematics, appearing in fields as diverse as algebraic geometry \cite{bl,tata1}, number theory \cite{tata1, EichlerZagierJacobi1985}, differential geometry \cite{thetasurfaces}, integrable systems \cite{KricheverShiotaSoliton2013,SegurWaves2008}, discrete mathematics \cite{RegevStephensInequality2017}, cryptography \cite{fastgenus2} and statistics \cite{AgostiniAmendolaDiscreteGaussians2019}. 

We present a new package
\texttt{Theta.jl} for numerical computations of theta functions, programmed in Julia \cite{julia}. Our package is specialized for multiple
evaluations of theta functions for the same Riemann matrix $\tau\in\mathbb{H}_g$ and
different $z$, for small values of the genus $g$. Our implementation is based on
the algorithm from \cite{computing-theta}, which we extend to support computations of theta functions with
characteristics and derivatives of arbitrary order. Our package is
designed as an alternative to existing packages such as
\texttt{algcurves} \cite{computing-theta} in Maple,
\texttt{abelfunctions} \cite{sage-theta} in Sage and \cite{theta-matlab} in Matlab, with additional
functionalities and optimizations.

As an application, we study numerical approaches to the Schottky problem in genus five. The Schottky problem asks to recognize Jacobians of curves amongst
principally polarized abelian varieties, and is one of the central
questions in algebraic geometry since the 19th century \cite{GrushevskySchottkySurvey2012}. The first
nontrivial case of the Schottky problem is in genus four,
which is completely solved \cite{igusa}. For a recent approach linking computations and tropical geometry see \cite{CKS2018}. In this paper,
we describe computational approaches for studying the Schottky problem
in genus five, using our new package. In particular, we use \texttt{Theta.jl}
to compute the equations in \cite{FGSM, Accola} which give a weak
solution to the Schottky problem in genus five. We also use our
package for computations on the genus five Schottky problem for
Jacobians with a vanishing theta null, which is described in our companion paper \cite{schottkynull}. 

\medskip

\textbf{Acknowledgments:} We are grateful to Bernd Sturmfels for
suggesting to study the Schottky problem in genus five, and for his
continuous encouragement. We thank Paul Breiding, Gavril Farkas,
J\"org Frauendiener, Sam
Grushevsky, Christian Klein, Riccardo Salvati Manni, Andrey Soldatenkov, Sasha Timme and Sandro Verra for
useful comments and discussions. We thank the anonymous referees for their suggestions, which improved the quality of the paper and of the code. This project was initated at the Max
Planck Institute for Mathematics in the Sciences in Leipzig, which both authors would like to thank for the hospitality and support at various stages of this work. 

\section{Theta functions}
We recall here the basic definitions about theta functions with characteristics. For a more detailed account we refer to \cite{bl},\cite{tata1},\cite{igusa}. A
\emph{characteristic} is an element $m \in
(\mathbb{Z}/{2}\mathbb{Z})^{2g}$, which we represent as a
vector $m = \begin{bmatrix}\varepsilon \\\delta\end{bmatrix}$ where
  $\varepsilon,\delta \in \{0,1\}^g$. The \emph{Riemann theta function with characteristic} $m$ is defined as  
\begin{equation}\label{eqn:thetawithchar}
\theta[m](z,\tau) = \theta\begin{bmatrix}\varepsilon \\\delta\end{bmatrix}(z,\tau) = \sum_{n\in \mathbb{Z}^g} \mathbf{e}\left( \frac{1}{2}\left(n+\frac{\varepsilon}{2}\right)^t\tau \left(n+\frac{\varepsilon}{2}\right) + \left(n+\frac{\varepsilon}{2}\right)^t\left(z+\frac{\delta}{2}\right) \right)
\end{equation}
and it is a holomorphic function $\theta[m]\colon \mathbb{C}^g\times \mathbb{H}_g \to \mathbb{C}$.
The Riemann theta function in \eqref{eqn:theta} is a special
case of \eqref{eqn:thetawithchar}, where the characteristic is the all-zero vector.
The \emph{sign} of a characteristic $m$ is defined as $e(m) =
(-1)^{\varepsilon^t \delta}$, and we call a characteristic \emph{even} or
\emph{odd} if the sign is 1 or $-1$ respectively. As a function of
$z$, $\theta[m](z,\tau)$ is even (respectively odd) if
and only if the characteristic $m$ is even (respectively odd). There are $2^{g-1}(2^g+1)$ even theta characteristics and $2^{g-1}(2^g-1)$ odd theta characteristics.

The \emph{theta constants} are the functions on $\mathbb{H}_g$ obtained by evaluating the theta functions with characteristics at $z=0$,
\begin{equation}
\theta[m](\tau) = \theta[m](0,\tau)\,.
\end{equation} 
Theta constants corresponding to odd characteristics vanish identically.

\section{Numerically approximating theta functions} \label{section:approx}
We describe in this section the algorithm that we use to compute theta functions
in \texttt{Theta.jl}. In our implementation, we modify the algorithm from
\cite{computing-theta}, generalizing it for theta functions with
characteristics and derivatives of arbitrary order.

In this section, we separate
$z\in\mathbb{C}^g$ and $\tau\in\mathbb{H}_g$ into real and imaginary parts, by writing $z=x+iy$,
$\tau=X+iY$, where $x,y\in\mathbb{R}^g$ and $X,Y$ are real symmetric
$g\times g$ matrices. We also denote by $Y=T^tT$  the Cholesky decomposition of
$Y$, where $T$ is upper-triangular. For any real vector $V\in \mathbb{R}^g$, we use $[V]$ to denote the vector
whose entries are the entries of $V$ rounded to the closest integers,
and we denote $[\![V]\!] = V-[V]$. 

We set $v(n) = \sqrt{\pi}
T(n+[\![Y^{-1}y]\!])$ and we define the lattice $\Lambda =
\{v(n) \,|\, n\in\mathbb{Z}^g\}$, letting $\rho$ be the length of the
shortest nonzero vector in $\Lambda$. We denote by $\Gamma(z,x) = \int_x^{\infty} t^{z-1} e^{-t} dt$ the
incomplete Gamma function.

\subsection{Theta functions with characteristics}
In \cite{computing-theta}, Deconinck et al. derive numerical approximations of
the theta function and its first and second derivatives. We extend their results for computing theta
functions with characteristics and derivatives of arbitrary order.

We denote the $N$-th order derivative of the theta function along the
vectors $k^{(1)}, \ldots, k^{(N)}$ as 
\begin{align}
D\left(k^{(1)},\ldots, k^{(N)}\right)\theta(z,\tau) = \sum_{i_1,\ldots, i_N=1}^g
k_{i_1}^{(1)} \cdots k_{i_N}^{(N)} \frac{\partial^N
	\theta(z,\tau)}{\partial z_{i_1}\cdots \partial z_{i_N}}\,.
\end{align}

By the quasi-periodicity of the
theta function, it suffices to
consider inputs $z$ of the form $z=a+\tau b$, for $a,b\in [0,1)^g$.

\begin{theorem} \label{thm:uniformapprox-derivs-char}
	Fix $\tau\in\mathbb{H}_g$, $\epsilon >0$. Let  $k^{(1)},\ldots, k^{(N)}
	\in 
	\mathbb{C}^g$ be unit vectors, and let $R$ be the greater of $\frac{1}{2}\sqrt{g+2N+\sqrt{g^2+8N}} +
	\frac{\rho}{2}$ and the real positive solution of $R$ in 
	\begin{equation} \epsilon = (2\pi )^N
	\frac{g}{2}\left(\frac{2}{\rho}\right)^g  \sum_{j=0}^N \binom{N}{j}
	\frac{1}{\pi^{j/2}} \Vert T^{-1}\Vert ^j \sqrt{g}^{N-j} \Gamma\left(
	\frac{g+j}{2},
	\left(R-\frac{\rho}{2}\right)^2\right)\,. \end{equation}  
	For $z$ of
	the form $z=a+\tau b$,
	for $a,b\in [0,1)^g$, and ${\footnotesize\begin{bmatrix}\varepsilon \\\delta\end{bmatrix}}\in\{0,1\}^{2g}$, the $N$-th derivative $D(k^{(1)},\ldots, k^{(N)})\theta{\footnotesize\begin{bmatrix}\varepsilon
		\\\delta\end{bmatrix}}(z,\tau)$ of the theta function with characteristic is approximated by
	\begin{align}\label{eqn:uniformapprox-char-derivs-sum}
	\begin{split}
	& e^{\pi y^t Y^{-1} y} (2\pi i)^N \sum_{n\in C_{R}}\left(
	k^{(1)} \boldsymbol{\cdot} \left(n-\eta\right)\right) \cdots \left(k^{(N)}\boldsymbol{\cdot}
	\left(n-\eta\right)\right) \\
	&\times \mathbf{e}\left(\frac{1}{2} \left(n-\eta\right)^t X \left(n-\eta\right) +
	\left(n-\eta\right)^t \left(x+\frac{\delta}{2}\right)\right)
	e^{-\Vert v(n+\frac{\varepsilon}{2})\Vert ^2}\,,
	\end{split}
	\end{align}
	with an absolute error $\epsilon$ on the product of $(2\pi i)^N$
	with the sum, where $\eta = [Y^{-1}y]- \frac{\varepsilon}{2}$ and
	\begin{equation}\label{eqn:uniformapprox-char-u} 
	C_R = \{ n \in\mathbb{Z}^g\,|\, \pi (n-c)^t Y (n-c) < R^2\,,
	|c_j|<1\,,\forall j=1,\ldots,g\}\,. 
	\end{equation}
\end{theorem}

\begin{proof_of}{Theorem~\ref{thm:uniformapprox-derivs-char}}
	We first consider the case  $N=0$  without derivatives. Then the result for characteristics $\varepsilon=\delta=0$ is proven in {\cite[Theorem 2]{computing-theta}}, where they replace the deformed ellipsoid $C_R$ in \eqref{eqn:uniformapprox-char-u} with the ellipsoid
	\begin{equation} \label{eqn:pointwiseapprox-s}
	S_R = \{ n\in\mathbb{Z}^g\,|\, \Vert v(n)\Vert  < R\}\,. 
	\end{equation}
	For arbitrary characteristics $\varepsilon,\delta$, we see  from \eqref{eqn:thetawithchar} that we can compute the corresponding theta function
	in a similar way as the usual theta function, by translating $z$ to $z+\frac{\delta}{2}$, and
	translating the lattice points in the sum from $n$ to
	$n+\frac{\varepsilon}{2}$. Note that this only changes the real part of
	$z$, while the imaginary part stays the same. Hence the approximation in Theorem \ref{thm:uniformapprox-derivs-char} holds for theta functions with
	characteristics, if we  take the sum over the ellipsoid
	\begin{align} \label{eqn:pointwiseapprox-char-s}
	S_{R,\varepsilon} = \left\{ n\in\mathbb{Z}^g
	\,\bigg|\, \left\Vert v\left(n+\frac{\varepsilon}{2}\right)\right\Vert  < R\right\}\,.
	\end{align}
	To obtain a uniform approximation for any $z\in\mathbb{C}^g$ and any characteristic, we take the union of the ellipsoids
	$S_{R,\varepsilon}$ from \eqref{eqn:pointwiseapprox-char-s} as
	$z$ and $\varepsilon$ vary. Since $v(n+\frac{\varepsilon}{2}) =
	\sqrt{\pi} T(n+[\![Y^{-1}y]\!] + \frac{\varepsilon}{2})$, and the entries
	of $[\![Y^{-1}y]\!] + \frac{\varepsilon}{2}$ have absolute value at most
	1, it follows that the deformed ellipsoid $C_R$ from \eqref{eqn:uniformapprox-char-u} is the union of the ellipsoids $S_{R,\varepsilon}$.
	
	To prove the result in the case of derivatives of order $N$, it will be enough to prove it for the case of zero characteristics, and then follow the same strategy as above. More precisely, we are going to prove the same statement as in \ref{thm:uniformapprox-derivs-char}, where $\varepsilon=\delta=0$ and $C_R$ is replaced by
	\begin{equation} \label{eqn:uniformapprox-u}
	U_R = \{ n \in\mathbb{Z}^g\,|\, \pi (n-c)^t Y (n-c) < R^2\,,
	|c_j|<1/2\,,\forall j=1,\ldots,g\}\,,
	\end{equation}
	To do so, we write the derivative as
	\[
	D(k^{(1)},\ldots, k^{(N)})\theta(z,\tau)
	=  (2\pi i)^N \sum_{n\in\mathbb{Z}^g} (k^{(1)}\boldsymbol{\cdot} n) \cdots (k^{(N)}\boldsymbol{\cdot} n) \mathbf{e} \left(\frac{1}{2}n^t \tau n + n^t z\right) 
	\]
	and then the error  in the approximation is 
	\begin{align*}
	\epsilon 
	&=  \bigg| (2\pi i)^N \sum_{n\in \mathbb{Z}^g \backslash U_R} \left(
	k^{(1)} \boldsymbol{\cdot} (n-[Y^{-1}y])\right) \cdots \left(k^{(N)}\boldsymbol{\cdot}
	(n-[Y^{-1}y])\right)  \\ 
	&\quad\quad\times\mathbf{e}\left(\frac{1}{2} (n-[Y^{-1}y])^t X (n-[Y^{-1}y]) +
	(n-[Y^{-1}y])^t x\right)e^{-\Vert v(n)\Vert ^2} \bigg|
	\end{align*}
	Since the $k^{(i)}$ have norm one, using the triangle inequality and the Cauchy-Schwartz inequality we can bound this by
	\begin{align*}
	\epsilon 
	& \leq (2\pi )^N   
	\sum_{n \in\mathbb{Z}^g \backslash U_R} \big\Vert n-[Y^{-1}y] \big\Vert ^N  e^{-\Vert v(n)\Vert ^2} 
	= (2\pi )^N 
	\sum_{n\in \mathbb{Z}^g \backslash U_R} \bigg\Vert  \frac{1}{\sqrt{\pi}} T^{-1} v(n) -
	Y^{-1} y \bigg\Vert ^N  e^{-\Vert v(n)\Vert ^2} 
	\end{align*}
	Using again the triangle inequality and the binomial expansion, we get to the bound
	\begin{align*}
	\epsilon &\leq  (2\pi )^N 
	\sum_{j=0}^N \binom{N}{j}
	\frac{1}{\pi^{j/2}} \Vert T^{-1}\Vert ^j \Vert Y^{-1}y\Vert ^{N-j} \sum_{n \in\mathbb{Z}^g \backslash U_R} \Vert v(n)\Vert ^j  e^{-\Vert v(n)\Vert ^2} 
	\end{align*}
	We then apply \cite[Lemma 2]{computing-theta} to get the bound
	\begin{align*}
	\epsilon 
	&\leq  (2\pi )^N 
	\sum_{j=0}^N \binom{N}{j}
	\frac{1}{\pi^{j/2}} \Vert T^{-1}\Vert ^j \Vert Y^{-1}y\Vert ^{N-j} \frac{g}{2}\left(\frac{2}{\rho}\right)^g \Gamma\left(
	\frac{g+j}{2}, \left(R-\frac{\rho}{2}\right)^2\right)\,, \\
	&\leq  (2\pi )^N
	\frac{g}{2}\left(\frac{2}{\rho}\right)^g 
	\sum_{j=0}^N \binom{N}{j}
	\frac{1}{\pi^{j/2}} \Vert T^{-1}\Vert ^j \Vert Y^{-1}y\Vert ^{N-j} \Gamma\left(
	\frac{g+j}{2}, \left(R-\frac{\rho}{2}\right)^2\right)\,.
	\end{align*}
	For inputs $z$ of the form $z=a+\tau b$, we can write $z$ as $z=a + (X+iY)b =
	(a+Xb) + iYb = x + iy$. Then $\Vert Y^{-1}y\Vert  = \Vert b\Vert  \leq
	\sqrt{g}$. Substituting this into the expression for $\epsilon$, the result follows.
\end{proof_of}

\begin{remark}
	The $R$ appearing in \ref{thm:uniformapprox-derivs-char} is computed numerically.
\end{remark}

\section{Computing theta functions in Julia}

\subsection{Interface} \label{section:julia-interface}
Our Julia package \texttt{Theta.jl} is available at the following website,
which has instructions and a link to more detailed documentation.

\begin{center}
\url{https://github.com/chualynn/Theta.jl}
\end{center}

We describe the basic interface of the package here. Starting with a matrix $\tau\in\mathbb{H}_g$, we first construct a
\texttt{RiemannMatrix} from it. This is a type in \texttt{Theta.jl} which
contains information needed to compute the theta function with input
$\tau$. As an example, we start with a genus 5 curve defined by the singular model
\begin{equation}
x^6y^2-4x^4y^2-2x^3y^3-2x^4y+2x^3y+4x^2y^2+3xy^3+y^4+4x^2y+2xy^2+x^2-4xy-2y^2-2x+1\,.
\end{equation}
We compute the Riemann matrix $\tau$ of the curve using the package
\cite{BruinSage2019} in Sage \cite{SAGE}, and we type it as an input
in Julia. We then construct a 
\texttt{RiemannMatrix} in \texttt{Theta.jl}, where we specify in the
input the options to compute a Siegel transformation, an error of
$10^{-12}$, and to compute derivatives up to the fourth order.\\

{\texttt{
julia> R = RiemannMatrix($\tau$, siegel=true, $\epsilon$=1.0e-12, nderivs=4);
}\\

We pick some input $z$ and compute the theta function
$\theta(z,\tau)$ as follows.

{\small
\begin{verbatim}
julia> z = [1.041+0.996im; 1.254+0.669im; 0.591+0.509im; -0.301+0.599im; 0.388+0.051im];
julia> theta(z, R)
-854877.6514446283 + 2.3935081163150463e6im
\end{verbatim}
}

We can compute derivatives of theta functions by specifying
the directions using the optional argument \texttt{derivs}. For instance, to compute $\frac{\partial^3
  \theta}{\partial z_3 \partial z_4} (z,\tau)$, we use

\begin{verbatim}
julia> theta(z, R, derivs=[[0,0,1,0,0], [0,0,0,1,0]])
1.0478325534969474e8 - 3.369999441122761e8im
\end{verbatim}

We can also compute derivatives of theta functions with
characteristics, where we specify the characteristic using the
optional argument \texttt{char}.

\begin{verbatim}
julia> theta(z, R, derivs=[[1,0,0,0,0]], char=[[0,1,0,0,1],[1,1,0,0,1]])
-2.448093122926732e7 + 3.582557740667034e7im
\end{verbatim}

\subsection{Algorithms} \label{subsection:thetajl-algo}
We describe here some details of the algorithms and the design choices
that we made in our implementation.

\subsubsection{Choice of ellipsoid}
We optimize our package for multiple evaluations of theta functions
at the same Riemann matrix $\tau$, and with different inputs
$z$, characteristics and derivatives. We do this using the approximation in Theorem~\ref{thm:uniformapprox-derivs-char}, which allows
us to compute derivatives of theta functions with characteristics,
for inputs $z$ of the form $z=a+\tau b$,
for $a,b\in [0,1)^g$. In this approximation, we take the sum over the
  deformed ellipsoid $C_{R}$ of \eqref{eqn:uniformapprox-char-u}, which
  depends only on the order $N$ of the derivative for a fixed
  $\tau$. Hence it suffices to compute the ellipsoids $C_R$ once for each order of the
  derivative that we are interested in, after
  which we can compute theta functions for any $N$-th order
  derivatives. These ellipsoids are stored in the
  \texttt{RiemannMatrix} type.

\subsubsection{Lattice reductions}\label{section:siegel}
In \cite{computing-theta}, the authors approximate the
length $\rho$ of the shortest vector of the lattice generated by $T$ using
the LLL algorithm by Lenstra, Lenstra and Lov\'asz \cite{lll}. This is a reasonable choice if $g$ is large, since the LLL algorithm gives a
polynomial time approximation, but with an error that grows exponentially with $g$.
In our implementation, since we focus on lattices with small
dimensions, we compute the shortest
vector exactly using the enumeration algorithm in
\cite{enum}. Moreover, by computing $\rho$ exactly, we obtain a
smaller ellipsoid \eqref{eqn:uniformapprox-char-u} than if we use the
LLL algorithm.

If we are interested in computing the theta function for a fixed $\tau$
at many values of $z$, it may be more efficient if we transform $\tau$
such that the ellipsoids in \eqref{eqn:uniformapprox-char-u} contain fewer lattice points. For this purpose, we use Siegel's algorithm, which iteratively finds a new matrix where the corresponding
ellipsoid has
a smaller eccentricity. In our implementation, we compute the Siegel transformation once for each Riemann
matrix, and work with the Siegel-transformed matrix for all computations. We use the algorithm for Siegel reduction described in \cite{computing-theta, theta-matlab}, where we use the algorithm
for HKZ reduction in \cite{hkz} as a subroutine.

\subsection{Comparisons with other packages}

The main advantage of \texttt{Theta.jl} over other packages
\cite{computing-theta, theta-matlab, sage-theta} is that we
support computations of theta functions with characteristics, as well as their derivatives, which to our knowledge is not
implemented elsewhere. Moreover, we make optimizations
described in Section~\ref{subsection:thetajl-algo} for faster
computations with a fixed Riemann matrix of low genus. 

We compare the performance of \texttt{Theta.jl} with the Sage package \texttt{abelfunctions} \cite{sage-theta}, by comparing the average time taken to compute the genus 5  FGSM relations of Section \ref{section:fgsm}, as well as to compute the
Hessian matrix of Section \ref{section:schottkynull}.  For our experiments, we sample matrices in the Siegel upper-half space
as follows. First we sample $5\times 5$
matrices $M_X, M_Y$ such that the entries are random floating point
numbers between $-1$ and $1$, using the random number generators in
Julia and NumPy. Then we sample $\tau\in\mathbb{H}_5$ as
$\tau=\frac{1}{2}(M_X+M_X^t) + M_Y^tM_Y i$. This is implemented in
\texttt{Theta.jl} for general dimensions $g$, in the function
\texttt{random\_siegel(g)}. In each experiment, we randomly sample
$1000$ such matrices, then we compute the FGSM relations and the
Hessian matrix using both packages on a standard laptop. We list in the table below the average time and standard
deviation.

\begin{center}
\begin{tabular}{|c|c|c|c|}
\hline
Experiment & Package & Average time (s) & Standard deviation (s) \\
\hline
FGSM & \texttt{Theta.jl} & 2.5 & 0.6 \\
& \texttt{abelfunctions} & 114.2 & 290.5 \\
\hline
Hessian & \texttt{Theta.jl} & 0.7 & 0.2\\
& \texttt{abelfunctions} & 20.3 & 58.0\\
\hline
\end{tabular}
\end{center}

One major reason for the faster runtime on
\texttt{Theta.jl} is the use of the Siegel transformation on the
Riemann matrix, which is not implemented in
\texttt{abelfunctions}. This also leads to the higher standard deviation in
the computations for the latter.

\section{Applications to the Schottky problem in genus five} \label{section:schottky}

Here we describe the main application that we had in mind when designing our package: experiments around the Schottky problem in genus five. We start with a brief account of the background of the problem, referring to \cite{GrushevskySchottkySurvey2012} for more details.

An abelian variety is a projective variety that has the structure of an algebraic group, and it is
a fundamental object in algebraic geometry. Especially important are principally polarized abelian
varieties, which can all be described in terms of Riemann matrices. For every $\tau \in \mathbb{H}_g$, we define the corresponding \emph{principally
polarized abelian variety (ppav)} as the quotient $A_{\tau} = \mathbb{C}^g / \Lambda_{\tau}$, where $\Lambda_{\tau} = \mathbb{Z}^g \oplus \tau\mathbb{Z}^g$ is a sublattice of $\mathbb{C}^g$. The polarization on $A_{\tau}$ is given by the \emph{theta divisor}
\begin{equation}
\Theta_{\tau} = \left\{  z \in A_{\tau} \,|\, \theta(z,\tau) = 0 \right\}\,.
\end{equation}

Two ppavs $A_{\tau}$ and $A_{\tau'}$ are isomorphic if and only if the corresponding Riemann matrices are related via an action of the symplectic group $\Gamma_g = \operatorname{Sp}(2g,\mathbb{Z})$. Hence, the quotient $ \mathcal{A}_g =  \mathbb{H}_g /\operatorname{Sp}(2g,\mathbb{Z}) $ is the \emph{moduli space of principally polarized abelian varieties of dimension $g$}. This is a quasi-projective variety of dimension $\dim \mathcal{A}_g = \dim \mathbb{H}_g = \frac{g(g+1)}{2}$, and we can look at the theta constants $\theta[m](0,\tau)$ as homogeneous coordinates on (a finite cover of) $\mathcal{A}_{g}$. 

Perhaps the most important example of abelian varieties are Jacobians of Riemann surfaces. For a Riemann surface $C$ of genus $g$, 
its \emph{Jacobian} is defined as the quotient
\begin{equation}
J(C) = H^0(C,\omega_C)^{\vee}/ H_1(C,\mathbb{Z})\,,
\end{equation}
where the lattice $H^1(C,\mathbb{Z})$ is embedded in $H^0(C,\omega_C)^{\vee}$ via the integration pairing
\begin{equation}\label{eqn:integrationofforms}
H^0(C,\omega_C)\times H^1(C,\mathbb{Z}) \longrightarrow \mathbb{C}\,, \qquad (\omega,\alpha) \mapsto \int_{\alpha} \omega\,.
\end{equation}
The Jacobian is a principally polarized abelian variety, and the corresponding Riemann matrix $\tau \in \mathcal{A}_g$ can be obtained by computing bases of $H^0(C,\omega_C)$ and $H^1(C,\mathbb{Z})$, as well as the integration pairing. 
This is implemented numerically in the packages \texttt{abelfunctions} \cite{sage-theta} and  \texttt{RiemannSurfaces} \cite{BruinSage2019} in Sage, and \texttt{algcurves} \cite{computing-theta} in Maple. 

The \emph{Schottky locus} $\mathcal{J}_g$ is the closure of the set of
Jacobian varieties in $\mathcal{A}_g$, and the \emph{Schottky problem}
asks for a characterization of $\mathcal{J}_g$ inside
$\mathcal{A}_g$. It is one of the most celebrated questions in
algebraic geometry, dating from the 19th century. There are
many possible interpretations of and solutions to the Schottky
problem. Here we focus on the most classical one, which asks for equations in the theta constants $\theta[m](0,\tau)$ that vanish exactly on the
Schottky locus. In this form, the Schottky problem is completely solved only in genus $4$, with an explicit equation given by Schottky and Igusa \cite{igusa}. A computational implementation and analysis of this solution was presented in \cite{CKS2018}.

The \emph{weak Schottky problem} asks for explicit equations that characterize Jacobians up to extra irreducible components. A solution to this problem was given in genus $5$ by Accola \cite{Accola}, and in a recent breakthrough, by Farkas, Grushevsky and Salvati Manni in all genera \cite{FGSM}. In the rest of this section, we discuss briefly these two solutions, together with related algorithms that we implemented in \texttt{Theta.jl}.  We also present a computational solution of a
 weak Schottky problem for genus five Jacobians with a theta null,
from our companion paper \cite{schottkynull}.

\subsection{Farkas, Grushevsky and Salvati Manni's solution}\label{section:fgsm}

In a recent preprint \cite{FGSM}, H. Farkas, Grushevsky and Salvati
Manni give a solution to the weak Schottky problem in arbitrary
genus. More precisely, for every genus $g\geq 4$ they give $\binom{g-2}{2} = \frac{(g-2)(g-3)}{2}$ explicit homogeneous equations of degree $2^{3\cdot 2^{g-4}+1}$ in the theta constants, such that their zero locus contains the Schottky locus as an irreducible component.

In the case of genus $5$, this gives $3$ equations of degree $128$. We implement them in the function \texttt{fgsm()} in
\texttt{Theta.jl}. Using the same example matrix $\tau$ from
Section~\ref{section:julia-interface}, the function \texttt{fgsm($\tau$)} gives us the output $7.850462293418876\mbox{e-}16$. This is expected
since $\tau$ is the Jacobian of a genus 5 curve.

\subsection{Accola's equations in genus 5}

A solution to the weak Schottky problem in genus $5$ was given already
by Accola \cite{Accola} in 1983, in the form of eight equations of degree $32$ in the theta constants whose zero locus contains the Schottky locus as an irreducible component. 
We implement these equations in the function
\texttt{accola()} in \texttt{Theta.jl}. Again using the example $\tau$ from
Section~\ref{section:julia-interface}, the function
\texttt{accola($\tau$)} gives us the output
$3.062334813867916\mbox{e-}9$, which is expected since $\tau$ is in
the Schottky locus.

\subsection{Schottky problem for Jacobians with a vanishing theta null}\label{section:schottkynull}

We describe here a variant of the Schottky problem focusing on
two-torsion points on Jacobians, referring to our companion article \cite{schottkynull} for a more complete account. A \emph{two-torsion point} on an abelian variety $A_\tau$ is a point $z\in A_{\tau}$ such that $2z=0$. These can be written as
\begin{equation}
z = \frac{\varepsilon}{2} + \tau \frac{\delta}{2}\,, \qquad \text{ for } \qquad  m = \begin{bmatrix}\varepsilon \\\delta\end{bmatrix} \in (\mathbb{Z}/2\mathbb{Z})^{2g}\,.
\end{equation}
Hence two-torsion points correspond to characteristics, and we say that such a point is \emph{even} or \emph{odd} if the corresponding characteristic is.  Observe that 
\begin{equation}
\theta\left( \frac{\varepsilon}{2} + \tau \frac{\delta}{2}, \tau \right) = 0 \qquad \text{ if and only if } \qquad \theta\begin{bmatrix}\varepsilon \\\delta\end{bmatrix}(0,\tau)=0 \,.
\end{equation}
Thus the two-torsion points in $\Theta_{\tau}$ correspond to the characteristics $m$ such that the theta constants $\theta[m](0,\tau)$ vanish. For this reason, we say that   
$A_{\tau}$ has a \emph{vanishing theta null} if it has an even two-torsion point in the theta divisor.   The abelian
varieties with this property have been intensely studied and they form a divisor $\theta_{\tnull}$ in $\mathcal{A}_g$.
The Jacobians with a vanishing theta null lie in the locus $\mathcal{J}_g\cap \theta_{\tnull}$ and they correspond to Riemann surfaces with an effective even theta characteristic. The Schottky problem in this case becomes that of recognizing $\mathcal{J}_g\cap \theta_{\tnull}$ inside $\theta_{\tnull}$.  

The first observation is that a vanishing theta null is automatically a singular point
of the theta divisor, because the partial derivatives $\frac{\partial\theta[m]}{\partial z_i}$ are odd. Hence
one is led to study the local structure of $\Theta_\tau$ around the
singular point, and the first natural invariant is the \emph{rank of
  the quadric tangent cone}, which corresponds to the rank of the
Hessian matrix of $\theta$ at the theta null. In particular, if a Jacobian has a vanishing theta null, then the quadric tangent cone has rank at most three. Hence
\begin{equation}
\mathcal{J}_g \cap \theta_{\tnull} \subseteq \theta^3_{\tnull}
\end{equation}
where we denote by $\theta^3_{\tnull}$ the locus of abelian varieties with a vanishing theta null whose quadric tangent cone has rank at most three.
Conversely, Grushevsky and Salvati Manni proved in
\cite{GrushevskySalvatiManniVanishing2008} that this inclusion is
actually an equality in genus $4$, confirming a conjecture of
H. Farkas. In the same paper, they ask whether $\mathcal{J}_g\cap
\theta_{\tnull}$ is an irreducible component of $\theta^3_{\tnull}$ in
higher genera, which would imply a solution to the weak Schottky problem for Jacobians with a vanishing theta null. The main result of our companion paper \cite{schottkynull} is an affirmative answer in genus $5$.

\begin{theorem}{\cite{schottkynull}}
	In genus five, the locus $\mathcal{J}_5\cap \theta_{\tnull}$ is an irreducible component of $\theta^3_{\tnull}$.
\end{theorem} 

We observe that the containment $\tau\in\theta^3_{\tnull}$ can be checked explicitly. Indeed, the condition of having an even two-torsion point in
the theta divisor can be checked by evaluating the finitely many theta
constants $\theta[m](0,\tau)$, and then numerically computing the rank of the
Hessian matrix. We present such a computation here, which is also in
\cite{schottkynull}. From the example in Section~\ref{section:julia-interface}, we use the
function \texttt{schottky\_null($\tau$)} in \texttt{Theta.jl}. The
output gives the even characteristic 

\begin{equation}
m = \begin{bmatrix} 1 & 0 & 0 & 1 & 0 \\ 1 & 0 & 1 & 1 & 0
 \end{bmatrix} 
\end{equation}
where the theta constant vanishes. The output also gives the
corresponding Hessian matrix

\begin{scriptsize}
\begin{equation*}
\begin{pmatrix}
 -2.79665+5.29764i & -9.57825-9.04671i  &  7.36305+2.28697i &
7.58338+5.34729i &   6.15667-1.90199i \\
-9.57825-9.04671i &  18.9738+8.34582i &   -23.1027-3.10545i &
-9.31944-0.822821i &  0.524289-3.64991i \\
7.36305+2.28697i & -23.1027-3.10545i  &  16.8441-1.15986i &
13.9363-4.56541i &  -3.32248+4.10698i \\
7.58338+5.34729i & -9.31944-0.822821i &  13.9363-4.56541i &
2.89309+1.21773i  &  3.86617-0.546202i \\
6.15667-1.90199i & 0.524289-3.64991i  & -3.32248+4.10698i &
3.86617-0.546202i & -12.9726-1.928i \\
\end{pmatrix}
\end{equation*}	
\end{scriptsize}
The Hessian has the eigenvalues
\begin{align*}
47.946229109152995 &+ 9.491932144035298i \\
-15.491689246713147 &+ 3.3401255907497958i \\
-9.512858919129267 &- 1.0587349322052013i \\
-2.7271385943272036\times 10^{-15} &- 1.1117459994936022i \times 10^{-14} \\
-5.698014266322794 \times 10^{-15} &+ 6.342925068807627i \times 10^{-15}
\end{align*}
so it has numerical rank 3 as expected.

\bibliographystyle{amsalpha}
\bibliography{ref}

\bigskip \bigskip  \bigskip

\noindent
\footnotesize 
{\bf Authors' addresses:}

\smallskip

\noindent Daniele Agostini,  MPl-MiS Leipzig,
\hfill  {\tt daniele.agostini@mis.mpg.de}

\noindent Lynn Chua, Caltech, 
\hfill {\tt lchua@caltech.edu}

\end{document}